\documentclass[aps,prb,reprint,superscriptaddress,amsmath,amssymb]{revtex4-2}

\usepackage{graphicx}
\usepackage{dcolumn}
\usepackage{bm}
\usepackage{color}
\usepackage{subcaption}
\usepackage{multirow}

\usepackage[utf8]{inputenc}
\usepackage{mathptmx}
\usepackage{etoolbox}
\usepackage[hidelinks]{hyperref}

\usepackage[version=3]{mhchem}
\usepackage[normalem]{ulem}

\DeclareUnicodeCharacter{0393}{\ensuremath{\Gamma}}
\DeclareUnicodeCharacter{03B1}{\ensuremath{\alpha}}
\DeclareUnicodeCharacter{03C0}{\ensuremath{\pi}}

\begin{document}

\author{Tammo van der Heide}
\affiliation{Bremen Center for Computational Materials Science, University of Bremen, 28359 Bremen, Germany}
\author{Ben Hourahine}
\affiliation{SUPA, Department of Physics, The University of Strathclyde, Glasgow, G4 0NG, United Kingdom}
\author{B\'alint Aradi}
\affiliation{Bremen Center for Computational Materials Science, University of Bremen, 28359 Bremen, Germany}

\author{Thomas Frauenheim}
\affiliation{Constructor University, School of Science, Campus Ring 1, Bremen, Germany}
\affiliation{Institute for Advanced Study, Chengdu University, Chengdu 610106, China}

\author{Thomas A.\ Niehaus}
\email{thomas.niehaus@univ-lyon1.fr}
\affiliation{Univ Lyon, Universit\'e Claude Bernard Lyon 1, CNRS, Institut Lumi\`ere Mati\`ere, F-69622 Villeurbanne, France}

\title{Phonon-induced band gap renormalization by\\dielectric dependent global hybrid density functional tight-binding}

\date{\today}

\begin{abstract}
Global hybrid exchange-correlation energy functionals within generalized Kohn-Sham density functional theory (GKS-DFT) have long been established as part of the standard repertoire for electronic structure calculations.
Accurate electronic bandstructures of solids are indispensable for a wide variety of applications and should provide a sound prediction of phonon-induced band gap renormalization at finite temperatures.
We employ our previously introduced formalism of general hybrid functionals within the approximate density functional method, DFTB, to present first insights into the accuracy of temperature dependent band gaps obtained by a dielectric-dependent global hybrid functional.
The work targets the prototypical group-IV semiconductors diamond and silicon.
Following Zacharias \emph{et al.}\ [Phys.\ Rev.\ Lett.\ 115, 177401 (2015)], we sample the nuclear wave function by stochastic Monte-Carlo integration as well as the deterministic one-shot procedure [Phys.\ Rev.\ B 94, 075125 (2016)] derived from it.
The computational efficiency of DFTB enables us to further compare these approaches, which fully take nuclear quantum effects into account, with classical Born-Oppenheimer molecular dynamic (BOMD) simulations.
While the quantum mechanical treatments of Zacharias \emph{et al.}\ yield band gaps in good agreement with experiment, calculations based on BOMD snapshots inadequately describe the renormalization effect at low temperatures.
We demonstrate the importance of properly incorporating nuclear quantum effects by adapting the stochastic approach to normal amplitudes that arise from the classical equipartition principle.
For low temperatures the results thus obtained closely resemble the BOMD predictions, while anharmonic effects become important beyond $500\,\mathrm{K}$.
Comparisons between DFTB parametrized from semi-local DFT, and global hybrid DFTB, suggest that Fock-type exchange systematically yields a slightly more pronounced electron-phonon interaction, hence stronger gap renormalization and zero-point corrections.
\end{abstract}

\maketitle

\section{Introduction}\label{sec:introduction}
Semiconductors form the backbone of modern electronic devices, playing a pivotal role in the development of technology that permeates various aspects of our daily lives.
As devices operate over a range of temperatures, the temperature dependence of the electronic band gap becomes a critical factor in ensuring stable and reliable operation.
Since the mid-20th century, various practitioners~\cite{history-1, history-2, history-3, varshni, history-4, history-5, history-6} have found that the band gap of traditional semiconductors like Ge, Si and GaAs decreases with temperature.
Subsequently, several empirical equations have been introduced and fitted to the experimental data~\cite{history-5}, with the Varshni equation~\cite{varshni} arguably being one of the most widely adopted empirical models.
Its accuracy has led to the redshift of the gap with temperature often being referred to as the Varshni effect.
For a more detailed historical review we refer to Ref.~\cite{epr-3}.
While most semiconductors obey Varshni's effect, materials with opposing behavior, like copper halides~\cite{inverse-varshni-1} and metal halide perovskites~\cite{inverse-varshni-2, inverse-varshni-3}, have been identified more recently.
The blueshift of the band gap with temperature is known as the inverse Varshni effect.

Usually, the Varshni effect is caused by an interplay of lattice expansion~\cite{latexp-0, latexp-1, latexp-2, latexp-3, latexp-4, latexp-5, latexp-6} and electron-phonon renormalization (EPR)~\cite{epr-1, epr-2, epr-3, epr-4, epr-5, epr-6}.
EPR refers to the impact of phonon-induced atomic vibrations on the electronic structure, encompassing zero-point motion renormalization (ZPR) and thermal vibration at finite temperatures.
In fact, the ZPR can be as large as $0.6\,\mathrm{eV}$~\cite{epr-2} and $0.37\,\mathrm{eV}$~\cite{dia-indirect-zpr-1, dia-indirect-zpr-2} for the direct and indirect band gap of diamond, respectively.

First-principles calculations of the optical properties of solids at finite temperature are usually based on either molecular dynamics (MD) simulations~\cite{md-1, md-2, md-3}, second-order perturbation theory (PT)~\cite{epr-1, epr-2, epr-3, epr-4, epr-5, pt-1, pt-2} or effective structures using a stochastic~\cite{mc} or one-shot~\cite{os} disorder approach based on the decomposition of the vibrational problem into normal modes.
The first of these approaches computes temperature-dependent band gaps as a time average of MD snapshots.
In practice, solids require computationally expensive large supercell calculations to sample the phonon wave-vector space, yet MD based approaches naturally include anharmonic vibrational modes beyond the harmonic approximation.
A limitation of classical MD is its neglect of nuclear quantum effects, spuriously yielding the classical Boltzmann statistics for phonons and ignoring zero-point motion.
However, path-integral molecular dynamic (PIMD) can recover the correct Bose-Einstein statistics and zero-point motion effects at the expense of computational efficiency~\cite{pimd-1, pimd-2, pimd-bosonic}.
In the second category, Allen-Heine-Cardona (AHC) theory~\cite{ahc}, which describes the thermal shift of the electronic energies on the basis of second-order perturbation theory within the adiabatic or non-adiabatic harmonic approximation, can be employed to obtain the temperature dependence of the electronic structure.
For instance, Giustino \emph{et al.}~\cite{epr-2} computed unperturbed band structures with the GW approximation~\cite{gw} and treated lattice dynamics at the density-functional perturbation level of theory~\cite{dfpt-review, dfpt-1, dfpt-2}.
The third type of approach generates effective structures with frozen phonon displacements, generated as a sum over Bose-Einstein distribution weighted normal modes~\cite{fp-1, fp-2, fp-3, fp-4, fp-5}.
A comprehensive review of the special displacement method is provided in Ref.~\cite{os-review}.
Most recently, Zacharias \emph{et al.}~\cite{os-anharmonic} extended the special displacement method to enable efficient calculations of temperature-dependent anharmonic phonon dispersions and electron-phonon couplings in strongly anharmonic systems, such as metal halide perovskites.

The density functional tight binding (DFTB)~\cite{dftb1, scc-dftb} method consistently applies carefully chosen approximations to the KS- and GKS-DFT~\cite{dft, ks-dft, gks-dft} energy functionals, providing relatively accurate schemes that can be applied to extended systems with requirements for large unit cells or long timescale molecular dynamics, which otherwise prevent treatment by higher-level methods.
Naturally, the shortcomings of (semi-)local DFT are inherited by KS-DFTB, e.g.\ resulting in a similar underestimation of band gaps for similar basis set sizes.
The recently introduced extension to general hybrid functionals for periodic systems~\cite{HYB-DFTB} enables us to gain first insights into the performance of dielectric-dependent global hybrid functionals within the DFTB method for describing the band gap of materials at finite temperatures.

In this paper, we parametrize (hybrid) DFTB for the prototypical non-polar covalent semiconductors with indirect band gaps, diamond and Si.
These parameters are then used to compare the stochastic and one-shot approaches of Zacharias \emph{et al.}\ with classical Born-Oppenheimer MD (BOMD) simulations.
The intention is to investigate the accuracy of temperature-dependent band gap predictions by a dielectric-dependent global hybrid functional at the DFTB level of theory, in particular the influence of exact exchange on electron-phonon renormalization.

This work is structured as follows:
In Sec.~\ref{sec:theory}, we briefly review the periodic hybrid DFTB formalism of Ref.~\cite{HYB-DFTB} and special displacement method of Zacharias \emph{et al.}, with a focus on the stochastic and one-shot approaches adopted throughout this work.
Section~\ref{sec:parameters} describes the process of generating (hybrid) DFTB parameters for the prototypical semiconductors diamond and Si, based on the popular PBEh~\cite{pbe0} functional, where the exact exchange fraction is chosen according to the inverse of the orientationally averaged and ion-clamped macroscopic dielectric constant $1 / \varepsilon_\mathrm{r}^\infty$.
Computational details are provided in Sec.~\ref{sec:comp-details}.
Section~\ref{sec:results} discusses the phonon-induced band gap renormalization obtained by the two special displacement methods, as well as classical BOMD simulations in an isothermal-isobaric (NPT) ensemble.
This leads into a discussion of the impact of incorporating proper nuclear quantum effects and anharmonicity at certain temperature regions.
For these group-IV semiconductors, our work indicates that (hybrid) DFTB is capable of qualitatively reproducing experimental data for the temperature-dependent band gap of solids, but that quantitative accuracy depends on the employed electronic parametrization.
We close with a summary of our findings and by providing a brief outlook in Sec.~\ref{sec:s&c}.

\section{Theory}\label{sec:theory}
All quantities are given in Hartree atomic units throughout.

\subsection{Periodic hybrid DFTB}\label{sec:theory-dftb}
The density functional tight binding (DFTB)~\cite{dftb1, scc-dftb} method fills the gap between first-principles electronic structure methods and conventional semi-empirical schemes.
While being two to three orders of magnitude faster than \emph{ab initio} DFT, it remains sufficiently accurate and paves the way to tackle problems usually considered to be out of reach for quantum mechanical atomistic simulations.

Self-consistent charge SCC-DFTB~\cite{scc-dftb} expands the Kohn-Sham total energy functional around a reference electron density $\rho_0$, up to second order in the density fluctuations $\delta \rho$:
\begin{align}\label{eq:taylor}
E[\rho_0+\delta \rho] &= E^{(0)}[\rho_0] + E^{(1)}[\rho_0,\delta \rho] + E^{(2)}[\rho_0,(\delta \rho)^2],
\end{align}
where $\rho_0$ is commonly constructed as a superposition of atomic densities of neutral spin-free atoms.

The zeroth order (so called repulsive) term solely depends on the reference density $\rho_0$, which is of key importance for the transferability of DFTB parameters to different chemical environments.
In practice, the repulsive term is commonly approximated as a sum of short-ranged atomic pair-potentials and fitted to an \emph{ab initio} reference~\cite{scc-dftb}.
Many-body corrections to this term have been recently addressed, for example by employing neural networks~\cite{tensor-rep, fortnet} or force-fields~\cite{chimes}.

In DFTB, Kohn-Sham orbitals are expanded into a small valence-only basis set $\{\phi_{\mu}(\mathbf{r})\}$ that arises from first principles calculations of neutral, spin-unpolarized pseudo-atoms~\cite{dftbbeginner}:
\begin{align}\label{eq:pseudo-atom}
\left[ -\frac{\Delta}{2} + v^\mathrm{eff}[\rho^\mathrm{atom}] + \left( \frac{r}{r_0}\right)^n \right]\phi_{\mu} &= \tilde{\varepsilon}_\mu \phi_{\mu}
\end{align}
In addition to the effective potential $v^\mathrm{eff}$ of KS-DFT, a confinement potential of power $n$ gives rise to free parameters of
the electronic structure part of the method, in particular the so-called compression radii $r_0$.
Note, that generally different confining radii are used for the density and wave function.

The Hamiltonian $\mathrm{H}$ and overlap $\mathrm{S}$ matrix elements are then evaluated in a two-center approximation and pre-tabulated for high-symmetry orbital configurations as a function of distance in Slater-Koster tables~\cite{sk-trans}.
Applying the Slater-Koster transformations to the coordinate system of the solid and solving the resulting generalized eigenvalue problem for periodic systems
\begin{align}\label{eq:dftb-eigenproblem}
\sum_\nu c_{\nu i}(\mathbf{k})\left[ H_{\mu\nu}(\mathbf{k}) - \varepsilon_i S_{\mu\nu}(\mathbf{k}) \right] = 0,\quad \forall i
\end{align}
self-consistently, yields eigenvector coefficients $c_{\nu i}(\mathbf{k})$ and eigenvalues $\varepsilon_i$ of eigenstate $i$, with orbital indices $\mu,\nu$.

Incorporating Fock-type exchange, as required for the hybrid exchange-correlation functionals employed in this work, leads to the additional energy contribution~\cite{HYB-DFTB}
\begin{align}\label{eq:E-kspace}
E^{x,\mathrm{CAM}} &= \frac{1}{2} \sum_{\mathbf{k}} w_{\mathbf{k}} \sum_{\mu\nu} \Delta H^{x,\mathrm{CAM}}_{\mu\nu}(\mathbf{k}) \Delta P_{\nu\mu}(\mathbf{k}),
\end{align}
with $k$-point weights $w_{\bm{k}}$ normalized as $\sum_{\bm{k}} w_{\bm{k}} = 1$, density fluctuations $\Delta P_{\nu\mu}(\mathbf{k})$ and Hamiltonian matrix elements
\begin{align}\label{eq:HFX-rspace}
\Delta H^{x,\mathrm{CAM}}_{\mu\nu}(\mathbf{k}) &= -\frac{1}{8} \sum_{\lambda\kappa} \sum_{\mathbf{g}\mathbf{h}\mathbf{l}} \Delta P_{\lambda\kappa}(\mathbf{g} + \mathbf{h} - \mathbf{l}) S_{\lambda\mu}(\mathbf{h}) S_{\kappa\nu}(\mathbf{l}) \notag
\\
&\times \Big[ \gamma_{\mu\nu}^\mathrm{CAM,HF}(\mathbf{g}) + \gamma_{\mu\kappa}^\mathrm{CAM,HF}(\mathbf{g} - \mathbf{l}) \notag
\\
&+ \gamma_{\lambda\nu}^\mathrm{CAM,HF}(\mathbf{g} + \mathbf{h}) + \gamma_{\lambda\kappa}^\mathrm{CAM,HF}(\mathbf{g} + \mathbf{h} - \mathbf{l}) \Big] \mathrm{e}^{-i\mathbf{k} \cdot \mathbf{g}}.
\end{align}
According to the Coulomb-attenuating method (CAM)~\cite{cam}, the parameters $\alpha, \beta$ and $\omega$ determine the fraction of global and long-range Fock-type exchange, as well as the value of the smooth range-separation function between the long- and short-range contributions.
For brevity, this parametric dependency of Eq.~\eqref{eq:HFX-rspace} is absorbed into the modified CAM $\gamma$-function $\gamma_{\mu\nu}^\mathrm{CAM,HF} = \alpha\gamma^\mathrm{fr,HF}_{\mu\nu} + \beta\gamma^\mathrm{lr,HF}_{\mu\nu}$ of DFTB.
The parametrizations $\gamma^\mathrm{fr,HF}_{\mu\nu}$ and $\gamma^\mathrm{lr,HF}_{\mu\nu}$ represent the Coulomb-type integrals with either the full- or long-range kernel, respectively, as defined in Eq.~(32) and (33) of Ref.~\cite{HYB-DFTB}.

A more comprehensive description of the DFTB method and its extensions is provided in Refs.~\cite{dftbbeginner, dftb+}.

\subsection{Special displacement method}\label{sec:mc-os}
In this section, we briefly recapitulate the stochastic~\cite{mc} and one-shot~\cite{os} approaches of Zacharias \emph{et al.}\ for
calculating temperature-dependent band gaps of solids.
From the theory of Williams~\cite{williams} and Lax~\cite{lax}, originally developed to study the vibrational broadening of the photoluminescence spectra of defects in solids, Zacharias \emph{et al.} derived an expression for the imaginary part of the dielectric function at finite temperature $T$.
This approach yields the optical spectrum over the full frequency range and can be directly compared to experimental data.
We are specifically interested in the temperature-dependent band gap $E_\mathrm{g}(T)$ and analogously write
\begin{align}\label{eq:wl-egap}
E_\mathrm{g}(T) = \frac{1}{Z} \sum_{n} \mathrm{e}^{-\frac{E_n}{k_\mathrm{B} T}} \langle E_\mathrm{g}(x) \rangle_n,
\end{align}
with the canonical partition function $Z = \sum_{n} \exp[ -E_n / (k_\mathrm{B} T)]$, energy of a nuclear quantum state $E_n$ in the Born-Oppenheimer approximation and Boltzmann constant $k_\mathrm{B}$.
We evaluate the band gap $E_\mathrm{g}(x)$ of a particular configuration $x$ of all atomic coordinates with clamped nuclei.
To this end, large supercells (see below) of the target system are simulated at the $\Gamma$-point, and the band gap is taken to be the smallest energy difference between occupied and unoccupied bands.
This protocol automatically provides the indirect band gap in relevant systems.

Each expectation value $\langle\cdot\rangle_n$ is with respect to the $n^\mathrm{th}$ nuclear quantum state.
We follow Zacharias \emph{et al.}, evaluate Eq.\,\eqref{eq:wl-egap} in the harmonic approximation and apply Mehler's formula~\cite{mehler, mehler-engl}, yielding
\begin{align}\label{eq:tdep-gap}
E_\mathrm{g}(T) = \prod_{\nu} \int \frac{\exp \big[-x_\nu^2 / (2 \sigma_{\nu,T}^2) \big]}{\sqrt{2\pi} \sigma_{\nu,T}} E_\mathrm{g}(x)\,d x_\nu.
\end{align}
The harmonic oscillator wave function $\exp[-x_\nu^2 / (2 \sigma_{\nu,T}^2)] / (\sqrt{2\pi} \sigma_{\nu,T})$ contains Gaussian widths $\sigma_{\nu,T}$ associated with the $\nu^\mathrm{th}$ normal coordinate $x_\nu$, Bose-Einstein occupations $n_{\nu,T}$ and zero-point vibrational amplitudes $l_\nu$:
\begin{align}
n_{\nu,T} &= (\exp[ \Omega_\nu / (k_\mathrm{B} T) ] - 1)^{-1},
\\
l_\nu &= (2 M_\mathrm{p} \Omega_\nu)^{-\frac{1}{2}},
\\
\sigma_{\nu,T} &= \sqrt{2 n_{\nu,T} + 1}l_\nu.
\end{align}
Here, $\Omega_\nu$ denotes the angular vibrational frequency of the $\nu^\mathrm{th}$ normal mode, i.e.\ harmonic oscillator, and in line with Refs.~\cite{mc, os} we choose the proton mass $M_\mathrm{p}$ to be our reference mass.
The stochastic approach to Eq.\,\eqref{eq:tdep-gap} employs importance-sampled Monte Carlo integration~\cite{fp-2} and averages over multiple atomic configurations with displacements
\begin{align}\label{eq:normal-to-displ}
\Delta R_{\kappa\alpha} &= \sqrt{\frac{M_\mathrm{p}}{M_\kappa}} \sum_\nu e_{\kappa\alpha,\nu} x_\nu,
\end{align}
of atom $\kappa$ with mass $M_\kappa$, along the Cartesian direction $\alpha$.
Diagonalizing the dynamical matrix of the system yields (gauge-corrected) eigenmodes $e_{\kappa\alpha,\nu}$ in ascending order with respect to their eigenfrequencies $\Omega_\nu$.
We can find a unique back-transformation to Eq.\,\eqref{eq:normal-to-displ} by exploiting the orthogonality of the rows and columns of the eigenvectors:
\begin{align}\label{eq:displ-to-normal}
x_{\nu} &= \frac{1}{\sqrt{M_\mathrm{p}}} \sum_{\kappa = 1}^{N} \sqrt{M_\kappa} \sum_{\alpha = 1}^3 \Delta R_{\kappa\alpha} e_{\kappa\alpha,\nu}.
\end{align}
For a supercell of $N$ atoms, the normal amplitudes $x_\nu$ of Eq.\,\eqref{eq:normal-to-displ} are then generated from a set of $3N - 3$ (i.e.\ translational modes removed) random numbers $0 < t < 1$
\begin{align}\label{eq:mc-normalcoords}
x_\nu &= \sqrt{2} \sigma_{\nu,T} \mathrm{erf}^{-1}(2t - 1),
\end{align}
where $\mathrm{erf}^{-1}$ refers to the inverse error function.

In Ref.~\cite{os}, Zacharias \emph{et al.}\ provide a formal proof that, in the limit of large supercells, a single atomic configuration is sufficient to evaluate Eq.\,\eqref{eq:tdep-gap}.
To make a clear distinction, we denote the atomic displacements of this deterministic one-shot method by $\Delta \tau_{\kappa\alpha}$:
\begin{align}\label{eq:os}
\Delta \tau_{\kappa\alpha} &= \sqrt{\frac{M_\mathrm{p}}{M_\kappa}} \sum_\nu (-1)^{\nu - 1} e_{\kappa\alpha,\nu} \sigma_{\nu,T}.
\end{align}
Note, that the different eigenmodes in Eq.\,\eqref{eq:os} contribute with alternating signs.
An improved numerical evaluation of Eq.\,\eqref{eq:tdep-gap} at fixed supercell size can be achieved by considering a hierarchy of atomic configurations, generated by iteratively partitioning the set of modes in two halves and applying a sign swap on one of the resulting subsets~\cite{os}.
Below, we will analyze the accuracy of the one-shot method and investigate the convergence properties of the hierarchical approach towards the exact result.

Williams-Lax (WL) theory treats the nuclei quantum mechanically and deals with their statistics in a consistent manner.
In order to rationalize results from MD simulations, we experimented with a hybrid WL scheme that is based on classical statistics.
According to the equipartition principle (EQP), in thermal equilibrium each normal mode has an average total energy of $k_\mathrm{B} T$, with exactly half of the energy assigned to the kinetic and potential terms.
By expressing the potential energy of a normal mode via the spectral decomposition of the dynamical matrix, it follows that
\begin{align}
\langle x_\nu^2 \rangle_T^\mathrm{eqp} &= \frac{k_\mathrm{B} T}{M_\mathrm{p}}\frac{1}{\Omega_\nu^2}.
\end{align}

Performing the Monte-Carlo integration of Eq.\,\eqref{eq:tdep-gap} with the stochastic normal amplitudes
\begin{align}\label{eq:classical-normal}
x_\nu^\mathrm{eqp} &= \sqrt{2} \frac{\sqrt{k_\mathrm{B} T /M_\mathrm{p}}}{\Omega_\nu} \mathrm{erf}^{-1}(2t - 1),
\end{align}
analogously to Eq.\,\eqref{eq:mc-normalcoords}, gives rise to an approach that we term MC(eqp) in the following.
In the limit of high temperatures, Eq.\,\eqref{eq:mc-normalcoords}, that fully takes nuclear quantum effects into account and yields Bose-Einstein statistics for phonons, approaches Eq.\,\eqref{eq:classical-normal}.

\section{Hybrid DFTB parameters}\label{sec:parameters}
The electronic parameters, established for the purpose of carrying out the investigations in Sec.~\ref{sec:results} of this work, have been generated using the SkProgs~\cite{skprogs-github} parametrization suite.

The web repository \url{www.dftb.org} is the primary address for obtaining Slater-Koster sets, which have been generated by the DFTB community.
Although general purpose parametrizations of carbon and silicon exist, with prime examples being the matsci-0-3~\cite{matsci-0-3-1, matsci-0-3-2} and pbc-0-3~\cite{pbc-0-3} set for solids in materials science, they all share certain limitations.
Restrictions arise from resorting to a minimal basis, generally leading to a poor representation of conduction bands in particular, compatibility pitfalls due to the choice of superposition strategy~\cite{scc-dftb} (potential vs.\ density superposition) or employing superseded exchange-correlation density functionals.
Arguably, the siband-1-1~\cite{siband-1-1} parameters pose an exception in the sense that they produce an accurate silicon bandstructure.
However, siband-1-1 is fitted to experimental reference data, which emphasizes the demand for parameters that are consistently fitted to the Hamiltonian that DFTB strives to resemble, which is in fact DFT.

To address this, we re-parametrized the elements carbon and silicon following a semi-automatic approach, while focusing on their respective diamond structure ($Fd\overline{3}m$), denoted as C-dia and Si-dia throughout.
Both elements are represented by an extended basis that includes unoccupied $3d$ polarization orbitals.
The respective density compression radii match those of the pbc-0-3 set, as have been found to have a minor influence on the electronic bandstructure.
We generated Slater-Koster files (in density superposition mode) on a non-equidistant grid of wavefunction compression values, and performed an exhaustive grid search, revealing candidates yielding satisfactory bandstructures.
In order to quantify the agreement with first-principles DFT, a loss function is constructed that monitors the dispersion of the highest occupied and lowest unoccupied band.
DFT references are computed using the FHIaims~\cite{fhiaims, fhiaims-1, fhiaims-2} software package and numeric atom-centered orbitals as specified by the 'tight' species defaults.
The procedure is carried out for PBE-parametrized~\cite{gga-pbe} DFTB, consistently using the same functional for the reference calculations.
We assume that the resulting tuned electronic parameters, like compression radii and confinement power, hold for the dielectric-dependent global hybrid pendant DD-PBEh~\cite{pbe0} as well, which is an approximation to reduce the parametrization effort, establishing both semi-local and hybrid DFTB parametrizations.
For the latter, Tab.~\ref{tab:dielectric-constants} provides the employed fractions of Fock-type exchange, as obtained from the inverse orientationally averaged, ion-clamped, macroscopic dielectric constant of the two target materials. \par
\begin{table}[htbp]
\centering
\begin{tabular}{lccc}
\hline
Material & $\varepsilon_\mathrm{r}^\infty$ & $\alpha$ & Ref. \\ \hline
Si-dia   & 11.25                           & 0.089    & \cite{wot-srsh} \\
C-dia    & 5.55                            & 0.180    & \cite{wot-srsh} \\ \hline
\end{tabular}%
\caption{Orientationally averaged, ion-clamped, macroscopic dielectric constants and fractions $\alpha = 1 / \varepsilon_\mathrm{r}^\infty$ of Fock-type exchange used in the current work.}
\label{tab:dielectric-constants}
\end{table}\noindent
The final atomic compression radii, as listed in Tab.~\ref{tab:compression-radii}, are selected by hand from the set of candidates from the automatic grid search.
To prevent overfitting, we ensure reasonable transferability by further including the bandstructure of zincblende structured silicon carbide (SiC-zb) as part of the manual parameter selection process.
Comparisons to semi-local and hybrid DFT references are provided in Fig.~S2 of the Supplemental Material~\cite{supplmat}, indicating excellent transferability to the SiC-zb compound.
\begin{table}[htbp]
\centering
\begin{tabular}{llccclclc}
\hline
&  & \multicolumn{3}{c}{$r_0^\mathrm{wave}\,[a_0]$} &  & \multirow{2}{*}{$r_0^\mathrm{density}\,[a_0]$} &  & \multirow{2}{*}{$p^\mathrm{conf}$} \\ \cline{3-5}
Element &  & s     & p     & d     &  &       &  &   \\ \hline
C       &  & 3.50  & 5.00  & 2.35  &  & 7.0   &  & 2  \\
Si      &  & 3.75  & 3.75  & 4.25  &  & 6.7   &  & 10 \\ \hline
\end{tabular}%
\caption{Wavefunction and density compression radii as well as the power of the confinement potential entering Eq.\,\eqref{eq:pseudo-atom} of the newly established carbon and silicon parameters in Sec.~\ref{sec:parameters}.}
\label{tab:compression-radii}
\end{table}

Extending the basis of C and Si by unoccupied $3d$ polarization orbitals comes with one major challenge: the associated positive onsite energy of the unbound scattering-state is neither well-defined nor numerically stable for the isolated neutral pseudo-atom calculations of Eq.\,\eqref{eq:pseudo-atom}, which build the basis for generating Slater-Koster files.
We therefore treat the respective $3d$ onsite energy of C and Si as free, exchange-correlation functional dependent, parameters.
Consequently, in case of the material-specific dielectric-dependent DD-PBEh functional, the onsite energies are treated as material-specific parameters, whereas they solely depend on the element for the PBE functional.
Depending on the material, the ill-defined unoccupied onsite energies result in qualitatively incorrect bandstructures due to an incorrect
energetic ordering of states.
As a drastic example, an incorrect choice could lead to C-dia being metallic.
To address this pitfall, we again resort to a semi-automatic procedure that in a first step yields an estimate of the unoccupied onsite energies in the system by fitting the total density of states (DOS) of the respective material to its DFT reference.
Since DFTB is only expected to yield reliable results for a window of a few electronvolts around the Fermi level, the loss function is restricted to a range that includes the valence band maximum (VBM) and conduction band minimum (CBM).
We used a gradient-free procedure based on Bayesian optimization~\cite{bayesian-optimization} to iteratively optimize the unoccupied onsite energies.
A subsequent manual fine tuning resulted in the onsite energies of Tab.~\ref{tab:unoccupied-onsites}.
\begin{table}[htbp]
\centering
\begin{tabular}{llcc}
\hline
& & \multicolumn{2}{c}{$\varepsilon_\mathrm{d}$ [Ha]} \\
\cline{3-4}
Material &  & PBE       & DD-PBEh \\ \hline
C-dia    &  & 0.4415    & 0.4740  \\
Si-dia   &  & 0.0496    & 0.0909  \\ \hline
\end{tabular}%
\caption{Unoccupied $d$-orbital onsite energies in Hartree atomic units, as obtained from the DOS-fitting procedure and subsequent manual fine tuning outlined in Sec.~\ref{sec:parameters}.}
\label{tab:unoccupied-onsites}
\end{table}

Figure~S1 contains the electronic bandstructures of C-dia and Si-dia, calculated on the PBE- and DD-PBEh-parametrized DFTB levels of theory.
To compute the self-consistent density, we resorted to a Monkhorst-Pack~\cite{mp} $k$-point sampling of at least $13\times 13\times 13$ throughout, including for the respective DFT references.
Around the VBM, even minimal basis DFTB usually provides an accurate description of the bands and our parameters follow this trend as well.
As is well known, the extended basis leads to significant improvements in the conduction bands~\cite{extended-basis}.
These states are usually insufficiently described by a minimal basis, to an extent that for a minimal basis, both materials appear to have a direct band gap.
But, the basis set is far from being complete, generally leading to flatter bands in $k$-space, and a compromise between an accurate description around the direct and indirect band gaps is evident.
For Si-dia, the tuning procedure of the unoccupied onsite energy essentially shifts the conduction bands.
Since the loss function is restricted to an energy window that includes the CBM but not the direct gap of Si-dia, the parametrization exhibits good agreement in the region around the CBM, i.e.\ the indirect band gap.
However, this results in an underestimation of the direct gap as large as $1.3\,\mathrm{eV}$, afflicting both the semi-local and hybrid parametrization.
In contrast, the conduction band edge of C-dia is not only composed of the $3d$-orbitals, and hence the tuning procedure does not lead to flawless agreement of the indirect band gap with the DFT reference.
We find an overestimation of the indirect band gap of C-dia of about $0.8$ and $0.6\,\mathrm{eV}$ for the semi-local and hybrid parametrizations, respectively.
The vicinity of the $\mathrm{\Gamma}$-point, i.e.\ the direct band gap of C-dia, is reasonably well represented however.

\section{Computational details}\label{sec:comp-details}
All DFTB based calculations are performed using the DFTB+ software package~\cite{dftb+}.
Born-Oppenheimer molecular dynamic simulations are computed in an isothermal-isobaric (NPT) ensemble with a timestep of $1.0\,\mathrm{fs}$, using a Nos\'e-Hoover chain thermostat~\cite{nose-hoover-chain} and a Berendsen barostat~\cite{berendsen}.
For the thermostat, a coupling strength of $500\,\mathrm{cm}^{-1}$ is used, while the barostat operates isotropically at a pressure of $1\,\mathrm{atm}$ ($101325\,\mathrm{Pa}$) and $0.5\,\mathrm{ps}$ timescale.
Trajectories are computed for $5\times 5\times 5$ supercells of the conventional unit cell of C-dia and Si-dia (1000 atoms), using the pbc-0-3~\cite{pbc-0-3} parameters at the non-SCC-DFTB level of theory, without charge self-consistency.
We verified that the contributions of self-consistent iterations (as in SCC-DFTB) have only a negligible influence on the results, so that the additional computational effort can be omitted.
Newton's equation of motion is propagated over a total of $100\text{-}150\,\mathrm{ps}$ with the lattice constant extrapolated as part of an intermediate restart, in order to accelerate the pressure equilibration.
The extrapolation is based on the observation that the lattice constant approaches a temperature-dependent supremum and restarts the MD at an estimated upper bound calculated from fitting a limited growth function to a first segment of each trajectory.
Snapshot geometries are extracted randomly within the last $50\,\mathrm{ps}$ of the trajectories, with a minimal time separation of $10\,\mathrm{fs}$.
Based on the snapshots, we performed single-point calculations in the $\Gamma$-point approximation using the functionals PBE- and DD-PBEh-DFTB, employing the parameters of Sec.~\ref{sec:parameters}.
Convergence of the band gap with supercell size is depicted in Fig.~S3.
We compute band gaps as a time average of $500$ snapshot calculations for each temperature that is considered.
The standard deviation provides an estimate of the amount of variation between the individual snapshots.

The fitness of the pbc-0-3 parameters for accurately describing phonon-induced phenomena is validated by calculating the phonon bandstructures of C-dia and Si-dia using the Phonopy~\cite{phonopy-1, phonopy-2} code.
We used $8\times 8\times 8$ supercells built from the respective primitive unit cell (1024 atoms) and performed single-point calculations at the $\Gamma$-point.
Figure~S4 illustrates the resulting phonon bandstructures, indicating reasonable agreement with experimental references in line with the conclusions of Ref.~\cite{dftb-4-phonons}.

For the special displacement based approaches, we relax pristine diamond and silicon bulk and build a $5\times 5\times 5$ supercell of the respective conventional unit cell.
We solve the vibrational problem on the basis of the pbc-0-3 parameters, by first calculating the Hessian matrix of the supercells using DFTB+ and a finite-difference step of $10^{-4}\,a_0$.
Subsequently, the dynamical matrix is constructed and diagonalized at $\mathbf{q} = 0$, using the MODES code (part of the DFTB+ software package).
A first post-processing step involves the sorting of eigenfrequencies and corresponding eigenvectors in ascending order.
We also fix the gauge of each vibrational mode by choosing the sign of each eigenvector so that its first nonzero element is positive.
Translational modes with zero frequency are omitted in each sum over normal modes.
The $3N - 3$ pseudo-random numbers $\{t\}$ of Eq.\,\eqref{eq:mc-normalcoords} and Eq.\,\eqref{eq:classical-normal} are generated from scrambled Sobol'~\cite{sobol, sobol-implementation} numbers, as implemented by the SciPy project~\cite{sobol-scrambling, scipy}.
Atomic masses are taken from the pbc-0-3 parameters, that assume isotope averages weighted by natural abundance.
We perform $\Gamma$-point single-point calculations on all geometries generated by either of the special displacement methods.
To ensure comparability between all methods, in terms of the underlying lattice expansion, the equilibrated lattice constants obtained as a time average over the $500$ MD snapshots mentioned above are assumed throughout.
All lattice constants are listed in Tab.~S1.
Figure~\ref{fig:MD-lattice-expansion} further illustrates the values of Tab.~S1 and provides experimental and semi-empirical references, as discussed later in Sec.~\ref{sec:results}.

\section{Results and discussion}\label{sec:results}
In the following, we discuss the merits and drawbacks of several methods to account for EPR effects within the DFTB formalism.
Naturally, different temperature regions impose specific demands on the underlying theory.
While an adequate description at low temperatures strongly depends on nuclear quantum effects, correct Bose-Einstein statistics for phonons in particular, anharmonicity may be negligible for the relatively rigid covalent materials considered in this work.
In contrast, the high-temperature limit might be influenced by sizable anharmonic contributions, whereas the phonon occupation approaches classical equipartition and nuclear quantum effects are diminishing.

The methods under consideration include MC integration and one-shot evaluation (OS) of Eq.\,\eqref{eq:tdep-gap}, Born-Oppenheimer molecular dynamic simulations (MD) and a modified MC scheme, that replaces the quantum normal amplitudes of Eq.\,\eqref{eq:mc-normalcoords} with their classical distribution of Eq.\,\eqref{eq:classical-normal} derived from the EQP.
Since the MC and OS schemes assume Bose-Einstein statistics for phonons, they are expected to provide an adequate description at moderate temperatures, including ZPR.
However, MC and OS are formulated in terms of the harmonic approximation and therefore lack any anharmonic contributions.
MD simulations, in contrast, neglect nuclear quantum effects such as ZPR, but consider anharmonicity.
MC(eqp) takes neither of these effects into account and constitutes a valuable tool in order to rationalize the differences between the
various methods.

Table~\ref{tab:zpr-mc-5x5x5} lists the indirect band gap ZPR of C-dia and Si-dia at the PBE- and DD-PBEh-DFTB levels of theory.
In fact, obtaining reliable estimates of ZP corrections from experimental data is highly non-trivial.
\citeauthor{C-dia_zpr_410}~\cite{C-dia_zpr_410} studied the sensitivity towards the employed extrapolation scheme and found that the same experimental data of the indirect gap of C-dia taken from Ref.~\cite{dia-indirect-zpr-1} yields ZP corrections ranging from $\text{-}290$ to $\text{-}510\,\mathrm{meV}$, depending on the extrapolation scheme.
The most accurate result is obtained by a 4$^\mathrm{th}$-order phonon dispersion model, which suggests an experimental ZPR of $\text{-}410\,\mathrm{meV}$~\cite{C-dia_zpr_410}.
Somewhat older experimental estimates of $\text{-}340$~\cite{C-dia_zpr_340} and $\text{-}370\,\mathrm{meV}$~\cite{dia-indirect-zpr-2} are in good agreement with \emph{ab initio} LDA-DFT data, among others, $\text{-}330$~\cite{epr-4}, $\text{-}334$~\cite{C-dia_zpr_334}, $\text{-}343$~\cite{C-dia_zpr_343}, $\text{-}344$~\cite{C-dia_zpr_344} and $\text{-}345\,\mathrm{meV}$~\cite{os}.
Our results for C-dia tend to overestimate the ZPR effect compared to most of the references, yet both the semi-local and hybrid DFTB value fall within the uncertainty range determined in Ref.~\cite{C-dia_zpr_410}.
Comparing semi-local and hybrid DFTB suggests that Fock-type exchange admixed to the density functional approximation (DFA) systematically predicts a slightly stronger electron-phonon interaction and ZPR, following the trend of $GW$ quasiparticle corrections~\cite{fp-5, GW-corr-4-zpr}.
However, while incorporating Fock-type exchange induces an increase of the high-temperature slope of the indirect band gap of C-dia by a few percent, it falls short the approximately $40\%$ found by \citeauthor{fp-5}\cite{fp-5} for the direct gap and many-body corrections to DFPT.
The ZPR of the indirect band gap of silicon has been computed at $56$~\cite{epr-4}, $57$~\cite{os}, $58$~\cite{C-dia_zpr_344} and $60\,\mathrm{meV}$~\cite{C-dia_zpr_334}.
Measurements report $62$~\cite{Si-dia_zpr_62_orig, epr-5} and $64\,\mathrm{meV}$~\cite{dia-indirect-zpr-2}.
Again, our results tend to overestimate the renormalization effect, which might be an artifact of not fully converged $k$- and $q$-point sampling as illustrated in Figs.~S3 and S6.
In particular, Fig.~S6 suggests that the ZPR of Si-dia actually converges to the mentioned references for larger supercell sizes.
\begin{table}[htbp]
\centering
\begin{tabular}{llcc}
\hline
& & \multicolumn{2}{c}{ZPR [meV]} \\
\cline{3-4}
Material &  & PBE-DFTB  & DD-PBEh-DFTB \\ \hline
C-dia    &  & -465      & -487         \\
Si-dia   &  & -78       & -90          \\ \hline
\end{tabular}%
\caption{Indirect band gap zero-point renormalization (ZPR) calculated at the PBE- and DD-PBEh-DFTB levels of theory, using the electronic parameters established in Sec.~\ref{sec:parameters}.
The stochastic Monte-Carlo evaluation of Eq.\,\eqref{eq:tdep-gap} is based on $5\times 5\times 5$ supercells of the conventional unit cell.}
\label{tab:zpr-mc-5x5x5}
\end{table}
\begin{figure*}[htbp]
\centering
\begin{subfigure}{0.5\textwidth}
  \centering
  \includegraphics[width=0.95\textwidth]{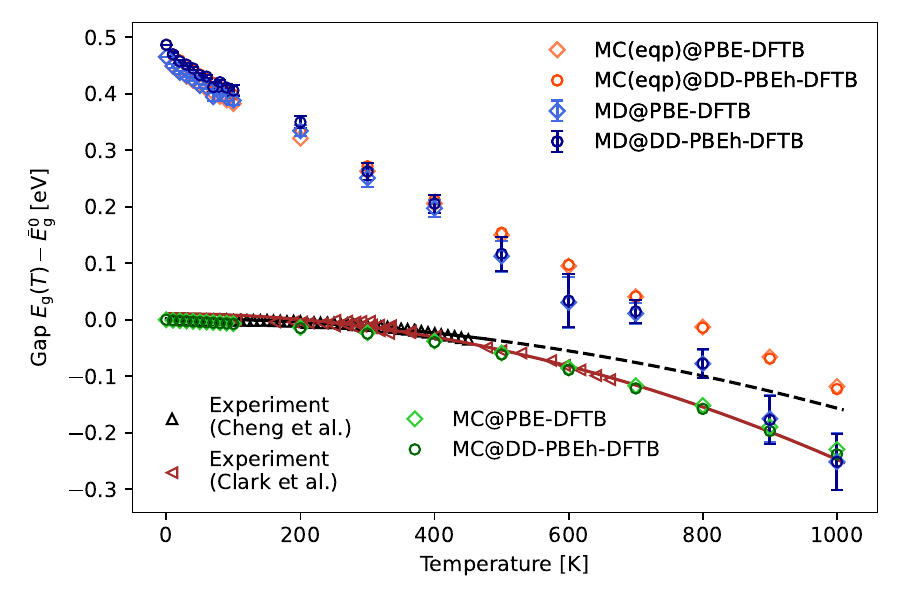}
  \caption{diamond}
  \label{fig:tdep-gap-diamond}
\end{subfigure}%
\hfill
\begin{subfigure}{.5\textwidth}
  \centering
  \includegraphics[width=0.95\textwidth]{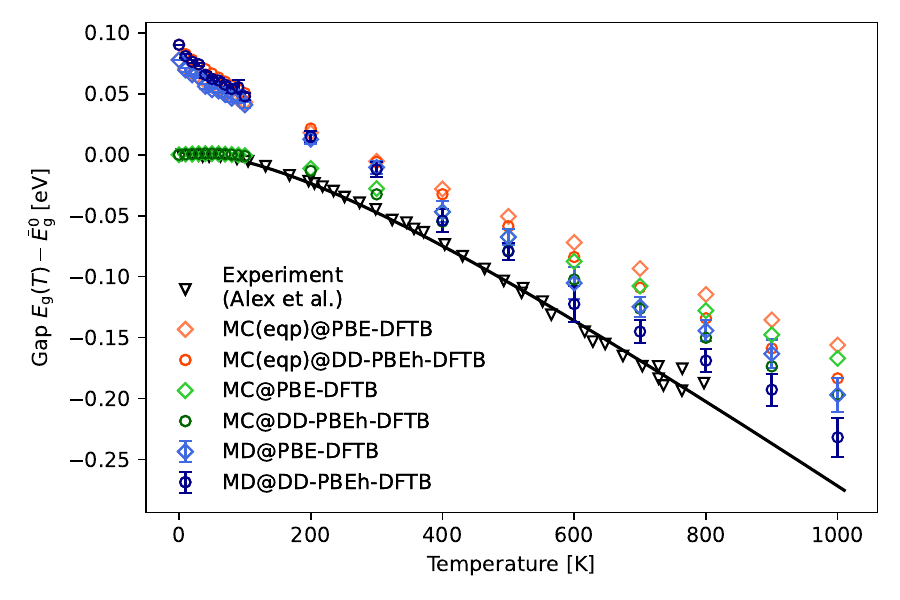}
  \caption{silicon}
  \label{fig:tdep-gap-silicon}
\end{subfigure}
\caption{Phonon-induced band gap renormalization calculated at the PBE- and DD-PBEh-DFTB levels of theory, using the electronic parameters established in Sec.~\ref{sec:parameters}.
Comparison is made between stochastic Monte-Carlo integration (MC), either based on normal amplitudes of Eq.\,\eqref{eq:mc-normalcoords} or Eq.\,\eqref{eq:classical-normal} and Born-Oppenheimer molecular dynamic simulations (MD).
Classical schemes are shifted by the respective zero-point renormalization ZPR(MC) obtained within the MC scheme, that is $\bar{E}^0_\mathrm{g} = E_\mathrm{g}(T=0\,\mathrm{K}) + \mathrm{ZPR(MC)}$ for MD and MC(eqp).
Solid lines refer to experimental references fitted by Varshni's equation to guide the eye and provide an extrapolation (see main text) to higher temperatures.
Further computational details are provided in Sec.~\ref{sec:comp-details}.
Experimental references are taken from Ref.~\cite{dia-exp-tdep-gap} (black triangles up), Ref.~\cite{dia-indirect-zpr-1} (brown triangles left) and Ref.~\cite{si-exp-tdep-gap} (black triangles down).}
\label{fig:tdep-gap}
\end{figure*}

Figure~\ref{fig:tdep-gap} shows the phonon-induced band gap renormalization for temperatures up to $1000\,\mathrm{K}$.
Experimental references are fitted by Varshni's equation to guide the eye and provide an extrapolation to higher temperatures, for which no
data points are available.
In the particular case of measurements by \citeauthor{dia-exp-tdep-gap} we depict the Varshni fit by a dashed segment.
This is to emphasize that the extrapolation has to be treated with caution, since the high temperature experimental data seems to indicate a different trend.
Given that absolute values of the band gap are strongly underestimated by the PBE functional (as evident from Fig.~S8), we compare the temperature induced changes instead.
Note that all methods without zero-point motion, i.e., MD and MC(eqp), are shifted by the respective ZPR obtained within the MC scheme.
There are clearly visible differences between the methods at lower temperatures.
Nuclear quantum effects, as captured by the special displacement methods, are neglected by classical MD simulations, leading to substantial errors at low temperatures for the latter.
Since our BOMD simulations do not yield the correct Bose-Einstein statistics for phonons, but rather compliance with the EQP, modes of higher energy are disproportionately excited with increasing temperature, leading to a distinct slope already apparent at low temperatures.
As demonstrated e.g.\ for diamond in Ref.~\cite{high-modes-influence}, it is precisely these higher-energy modes that exert the strongest influence on the alteration of the band gap.

In fact, Fig.~S7 provides evidence that for low temperatures our BOMD yields normal mode amplitudes in remarkable agreement with the equipartition theorem, indicating that the thermostat works as intended.
For high temperatures, however, this correspondence deteriorates noticeably due to the growing importance of anharmonicity.
It is therefore not surprising that we find an excellent agreement of the classical (harmonic) MC(eqp) scheme with BOMD at low temperatures,
suggesting that the discrepancy with respect to the quantum MC approach is predominantly rooted in the underlying phonon statistics.
BOMD is inherently inadequate to address phonon-induced gap renormalization at low temperatures and a proper treatment of nuclear quantum effects is indispensable.
Results exhibiting the same qualitative trends as the experimental references and thus Varshni's equation, are obtained only via the original MC method.
By comparing MC(eqp) with MD data, we further infer that anharmonic contributions become sizable when exceeding roughly $500\,\mathrm{K}$.
By about $1000\,\mathrm{K}$, anharmonic contributions for C-dia and Si-dia are as large as $0.1$ and $0.05\,\mathrm{eV}$, respectively, with important repercussions for the interpretation of the quantum MC data.
In principle, the MC results require a correction for these anharmonic contributions.
In light of this, the flawless agreement with the measurements of \citeauthor{dia-indirect-zpr-1} for C-dia at high temperatures appear accidental, anharmonicity corrected results would in fact overestimate the renormalization.
For Si-dia, in contrast, such a correction would yield even closer agreement with experiment.
In line with the observations made when computing the ZPR, Fock-type exchange admixed to the DFA predicts a slightly stronger gap renormalization.
However, the effect is small and somewhat concealed by the deviation between the methods.

\begin{figure*}[htbp]
\centering
\begin{subfigure}{0.5\textwidth}
  \centering
  \includegraphics[width=0.95\textwidth]{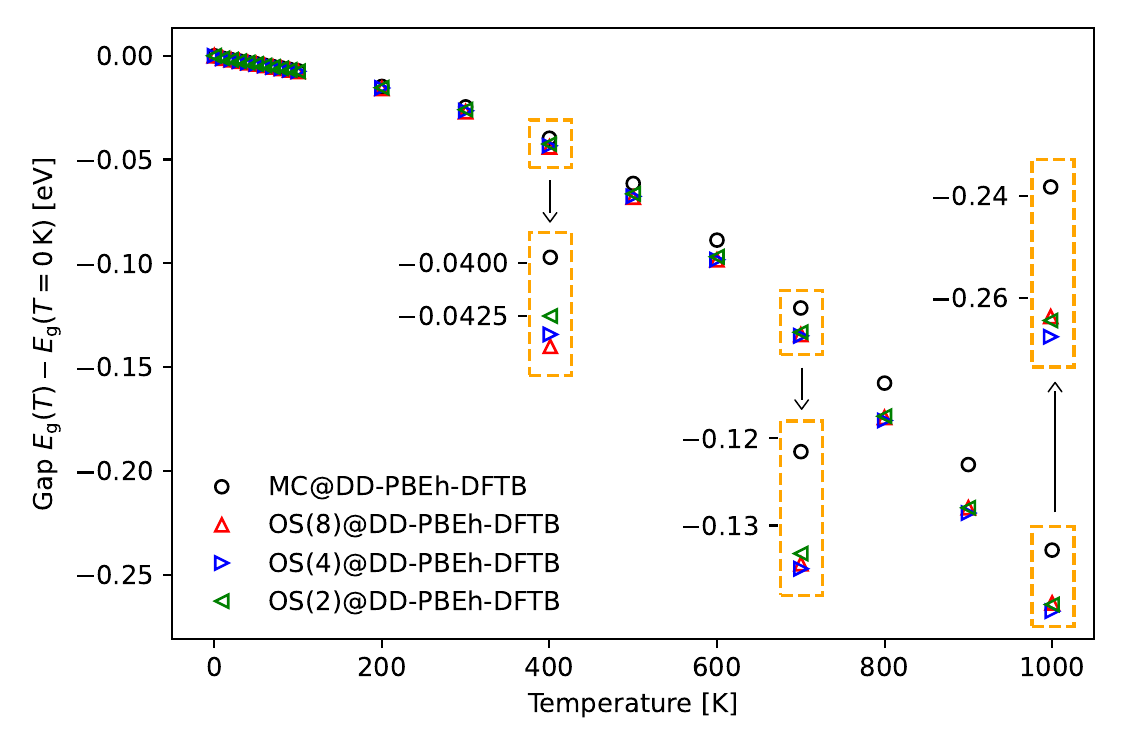}
  \caption{diamond}
  \label{fig:os-conv-diamond}
\end{subfigure}%
\hfill
\begin{subfigure}{.5\textwidth}
  \centering
  \includegraphics[width=0.95\textwidth]{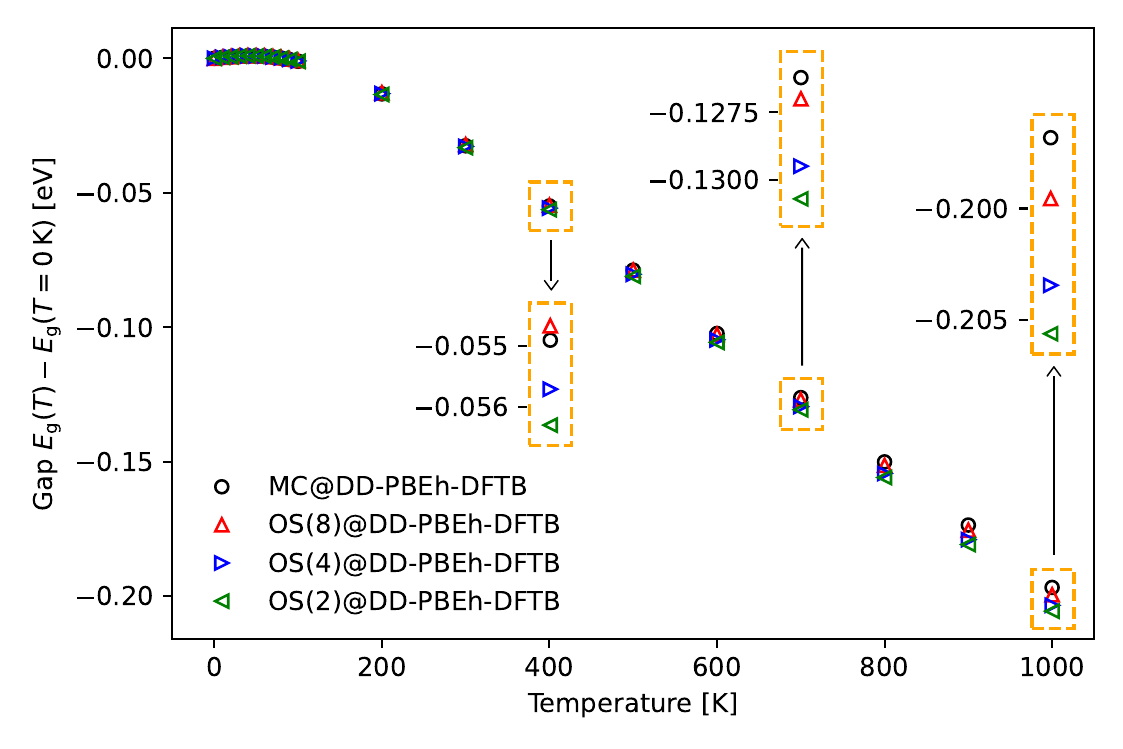}
  \caption{silicon}
  \label{fig:os-conv-silicon}
\end{subfigure}
\caption{Convergence analysis of the one-shot approach (OS) to Eq.\,\eqref{eq:tdep-gap}, with respect to the number of sign configurations (indicated in round brackets).
The phonon-induced band gap renormalization is calculated at the DD-PBEh-DFTB level of theory, using the parameters described in Sec.~\ref{sec:parameters}.
Further computational details are provided in Sec.~\ref{sec:comp-details}.}
\label{fig:os-conv}
\end{figure*}

During our investigation, we further observed that the one-shot approach does not exactly reproduce the converged Monte-Carlo integration of Eq.\,\eqref{eq:tdep-gap}, which we think requires further investigation.
In Ref.~\cite{os}, Zacharias \emph{et al.}\ proof that OS converges towards MC in the limit of large supercells and sufficient sign configurations.
Figure~\ref{fig:os-conv} shows the convergence of the OS approach with respect to  the number of sign configurations, in comparison with converged MC integration.
In case of Si-dia, the OS results systematically approach MC when increasing the number of sign configurations.
For C-dia the improvement is barely worth the additional computational effort and a significant discrepancy remains, even including eight sign configurations.
Since all of the MC based results shown in this work converged in the course of $<10$ random samples in line with Ref.~\cite{mc}, there is no point in iterating the sign hierarchy any further than eight configurations, as the OS approach has already lost its efficiency edge over the stochastic evaluation.
Therefore, we must note that although the OS method provides an extremely efficient and mostly accurate approximation to the fully converged solution, the inclusion of additional sign configurations is of limited use, at least for the systems considered in this work.
However, not every electronic structure method is inherently accurate enough for this slight discrepancy to actually matter.
The same findings emerge from Fig.~S9, that contains an equivalent study at the PBE-DFTB level of theory.

\begin{figure*}[htbp]
\centering
\begin{subfigure}{0.5\textwidth}
  \centering
  \includegraphics[width=0.95\textwidth]{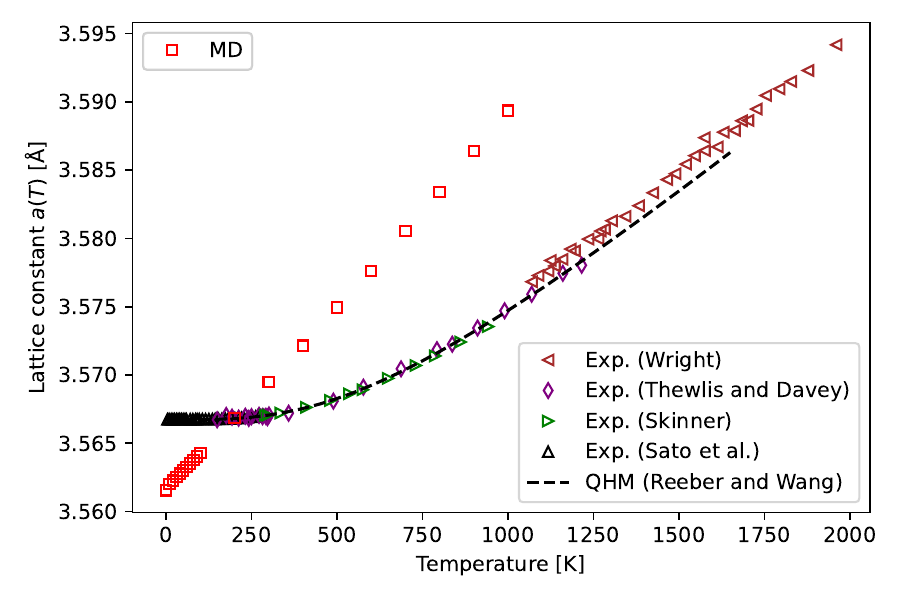}
  \caption{diamond}
  \label{fig:MD-lattice-expansion-diamond}
\end{subfigure}%
\hfill
\begin{subfigure}{.5\textwidth}
  \centering
  \includegraphics[width=0.95\textwidth]{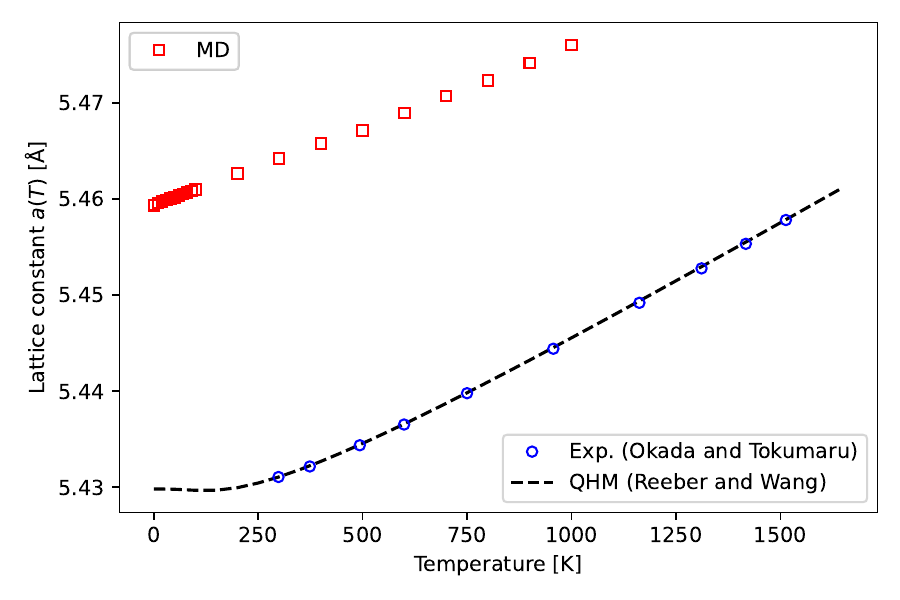}
  \caption{silicon}
  \label{fig:MD-lattice-expansion-silicon}
\end{subfigure}
\caption{Temperature-dependent lattice constants of diamond (C-dia) and silicon (Si-dia) bulk, computed as a time average over $500$ Born-Oppenheimer MD snapshots.
Snapshot geometries are extracted randomly within the last $50\,\mathrm{ps}$ of the trajectories, with a minimal time interval of $10\,\mathrm{fs}$.
At low temperatures, the qualitative discrepancy between BOMD simulations of this work and measurements arises from neglecting quantum nuclear effects, as discussed in Sec.~\ref{sec:results}.
Further computational details, including e.g.\ the thermo- and barostat settings of the isothermal-isobaric (NPT) ensemble, are provided in Sec.~\ref{sec:comp-details}.
Experimental references are taken from Ref.~\cite{latexp-exp-sato} (black triangles up), Ref.~\cite{latexp-exp-skinner} (green triangles right), Ref.~\cite{latexp-exp-thewlis_davey} (purple rhombs), Ref.~\cite{latexp-exp-wright} (brown triangles left) and Ref.~\cite{latexp-3} (blue circles), in addition to the semi-empirical quasi-harmonic model (QHM) of Reeber and Wang~\cite{latexp-2} (dashed line).}
\label{fig:MD-lattice-expansion}
\end{figure*}

We close with a brief discussion of the impact of thermal expansion.
Figure~\ref{fig:os-conv-diamond} highlights the slightly incorrect slope of the MC results at low temperatures, which is a consequence of the incorrect underlying temperature-dependent lattice constants extracted from equilibrated BOMD trajectories.
This is caused by the neglect of nuclear quantum effects, which leads to a qualitative distinction between measurements and classical MD, as illustrated by Fig.~\ref{fig:MD-lattice-expansion}.
The effect of pure lattice expansion is presented in Fig.~S5 for PBE- and DD-PBEh-DFT(B).
We find that the band gap of C-dia decreases with temperature, with opposing behavior in the case of Si-dia.
However, in comparison with the electron-phonon renormalization, the effect is small to an extent that it has been neglected in other works~\cite{mc, os}.

\section{Summary and Conclusion}\label{sec:s&c}
We have investigated phonon-induced band gap renormalization at the semi-local and dielectric-dependent global hybrid DFTB levels of theory, for the prototypical indirect semiconductors diamond and silicon.
The work is based on newly generated electronic DFTB parameters, following a general semi-automatic parametrization workflow.
It compares results obtained from a stochastic and one-shot approach to Williams-Lax (WL) theory with BOMD simulations.
To this end, the new parametrization is established for equilibrium structures calculated at the corresponding DFT level of theory, neither explicitly targeting application to EPR studies nor incorporating experimental data.
We find that DFTB provides qualitatively correct temperature-dependent band gap renormalizations.
Quantitative agreement with experiment, however, might be difficult to achieve using WL theory.
Zero-point renormalization (ZPR) estimates were still not fully converged for supercells containing roughly 2000 atoms.
Dielectric-dependent global hybrid DFTB, which admix Fock-type exchange with the density functional approximation, systematically yields slightly stronger electron-phonon interactions, including the ZPR.

A proper treatment of nuclear quantum effects is indispensable to obtain the same qualitative trends as seen in experiment.
BOMD grossly overestimates the renormalization at low temperatures, a consequence of the underlying classical phonon statistics.
We further studied the influence of nuclear quantum effects by employing a modified stochastic special displacement method based on normal amplitudes derived from EQP, that exhibits excellent agreement with classical BOMD calculations at low temperatures.
This hybrid method allowed to disentangle the contribution of quantum statistics and anharmonic corrections in a qualitative fashion.
We also performed a convergence analysis of the one-shot approach, revealing a small residual discrepancy with respect to fully converged Monte-Carlo integration that, especially in the case of diamond, remains even for up to eight sign configurations.
The comparison of gap values at different orders is therefore not a good indicator for convergence.
Still, the OS method is an extremely efficient and accurate method at temperatures below $500\,\mathrm{K}$.

Distinguished by its computational efficiency, DFTB allowed us to compare different theoretical approaches using a consistent electronic structure method over a wide temperature range.
CAM-DFTB retains full access to different treatments of the electron-electron interaction, e.g.\ it allows investigation of the influence of varying exchange-correlation functionals on the final result.
In the future, we envision application of path-integral MD with DFTB to incorporate both nuclear quantum and anharmonic effects in sufficiently large unit cells, a regime previously considered to be out of reach for first principles methods.

The data and Slater-Koster files that support the findings of this study are available within the paper and its Supplemental Material~\cite{supplmat}.

\begin{acknowledgments}
T.\ v.\,d.\,H.\ and T.\ A.\ N.\ acknowledge financial support from the German Research Foundation (DFG) through Grant No.\ FR2833/76-1 and FR2833/82-1, respectively.
The simulations were performed to similar shares on the HPC cluster \emph{Aether} at the University of Bremen, financed by the German Research Foundation (DFG) within the scope of Zukunftskonzept 66 “Ambitioniert und agil”, Bremen (GZ ZUK 66/1-2015) and the HPC cluster \emph{Lesum} at the University of Bremen, financed by the DFG (INST 144/506-1 FUGG).
\end{acknowledgments}


\bibliography{references}

\end{document}